# Paving the Way to Future Missions: the Roman Space Telescope Coronagraph Technology Demonstration

B. Mennesson[1], R. Juanola-Parramon[2], B. Nemati[3], G. Ruane[1], V. P. Bailey[1], M. Bolcar[2], S. Martin[1], N. Zimmerman[2], C. Stark[4], L. Pueyo[4], D. Benford[5], E. Cady[1], B. Crill[1], E. Douglas[6], B. S. Gaudi[7], J. Kasdin[8], B. Kern[1], J. Krist[1], J. Kruk[2], T. Luchik[1], B. Macintosh[9], A. Mandell[2], D. Mawet[10], J. McEnery[2], T. Meshkat[11], I. Poberezhskiy[1], J. Rhodes[1], A. J. Riggs[1], M. Turnbull[12], A. Roberge[2], F. Shi[1], N. Siegler[1], K. Stapelfeldt[1], M. Ygouf[1], R. Zellem[1] & F. Zhao[1]

[1] Jet Propulsion Laboratory, California Institute of Technology
[2] NASA Goddard Space Flight Center
[3] The University of Alabama in Huntsville
[4] Space Telescope Science Institute
[5] NASA Headquarters
[6] University of Arizona
[7] Ohio State University
[8] University of San Francisco
[9] Stanford University
[10] California Institute of Technology
[11] Caltech IPAC
[12] SETI Institute

Version 2 – September 2020

## Abstract

This document summarizes how far the Nancy Grace Roman Space Telescope Coronagraph Instrument (Roman CGI) will go toward demonstrating high-contrast imaging and spectroscopic requirements for potential future exoplanet direct imaging missions, illustrated by the HabEx and LUVOIR concepts. The assessment is made for two levels of assumed CGI performance: (i) current best estimate ("CBE") as of August 2020, based on laboratory results and realistic end-to-end simulations with JPL-standard Model Uncertainty Factors (MUFs); (ii) CGI design specifications inherited from Phase B requirements. We find that the predicted performance (CBE) of many CGI subsystems compares favorably with the needs of future missions, despite providing more modest point source detection limits than future missions. This is essentially due to the challenging pupil of the Roman Space Telescope; this pupil pushes the coronagraph masks' sensitivities to misalignments to be commensurate with future missions. In particular, CGI will demonstrate active low-order wavefront control and photon counting capabilities at levels of performance either higher than, or comparable to, the needs of future missions.

## 1. CGI top-level predicted performance and specifications vs. future missions' needs

Here we compare two parameterizations of Roman CGI performance, Current Best Estimates (CBEs) and design specifications (DSs), against the requirements for the HabEx and LUVOIR mission concepts. CGI CBEs are performance predictions based on analytic modeling, integrated modeling, and/or lab testing (Nemati et al. 2017, Krist et al. 2018, Zhou et al. 2019, Shi et al. 2019). CBEs include JPL-standard Model Uncertainty Factors (MUFs) when applicable. CBEs generated with MUFs are intended to provide performance predictions that have a high probability of being achieved in flight; actual performance may be somewhat higher in practice. Design specifications are the performance benchmarks against which CGI's preliminary design was evaluated. CGI has three baseline observing modes[1]: narrow field of view imaging at 575nm (NFOV), slit+prism spectroscopy at 730nm (SPEC), and wide field of view imaging at 825nm (WFOV). For each comparison, the most stringent design specification or CBE among the three CGI modes is adopted unless otherwise noted.

Tables 1 & 2 compare CGI CBEs as of August 2020 – as well as CGI design specifications – to the baseline requirements of the HabEx 4m off-axis telescope coronagraph (Gaudi et al. 2019), the LUVOIR A 15m on-axis telescope coronagraph and the LUVOIR B 8m off-axis telescope coronagraph (Fischer et al. 2019). For each coronagraph characteristic listed, the color coding indicates whether CGI estimated performance (or design specification) is better (green), in family with (blue) or significantly worse (red) than what is required for future missions. Table 1 concentrates on top-level characteristics that define a coronagraph discovery space – inner working angle (IWA) and point-source to star flux ratio detection limit at that IWA – as well as the coronagraph potential for spectral and polarimetric characterization. Table 2 concentrates on lower level technical requirements.

| Top Level Characteristics | CGI (CBEs) | CGI (Design Specifications) | HabEx Requirements | LUVOIR A/B Requirements |
|---|---|---|---|---|
| Inner Working Angle (in lambda/D) | 3<br>NFOV @ 575 nm<br>SPEC @ 730 nm | 4<br>NFOV @ 575 nm<br>SPEC @ 730 nm | 3.1<br>(@ 500 nm) | A: 4; B: 3.5<br>(@ 500 nm) |
| Flux Ratio Detection Limit at IWA | $10^{-8}$ (10σ) | $5 \times 10^{-8}$ (10σ) | $10^{-10}$ (10σ) | $10^{-10}$ (10σ) |
| Spectral Bandwidth | 10% (NFOV) - 15% (SPEC) | 10% (NFOV) - 15% (SPEC) | 20% | 20% |
| Spectral Resolution | 50 (SPEC) | 50 (SPEC) | 70 (VIS IFS) | 140 (VIS IFS) |
| Multiplanet Spectroscopy Capability | No | No | Yes | Yes |
| Polarimetric Capability | Yes<br>4 linear states, 2 at a time | Yes<br>4 linear states, 2 at a time | Yes | Yes |
| Entrance Pupil | 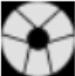 | 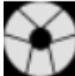 | 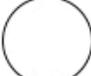 | 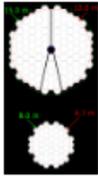 |
| Aperture type | on-axis 2.4m monolith<br>with 6 struts | on-axis 2.4m monolith<br>with 6 struts | off-axis 4m monolith | A: on-axis segmented with 13.5 m inscribed diameter<br>B: off-axis segmented with 6.7 m inscribed diameter |
| Primary Obscuration (linear in %) | 28% | 28% | 0 | A: 8% ; B: 0 |
| F/# | 1.3 | 1.3 | 2.5 | A: 1.3; B: 2.73 |

***Table 1****: Comparison of CGI top-level characteristics −current best estimates of performance (CBEs) as of August 2020 and design specifications − to the corresponding requirements for the HabEx and LUVOIR visible coronagraphs. For the CGI columns, the color-coding indicates whether CGI performance (CBE or design specification) is better (green), in family with (blue) or significantly worse (red) than what is required for future missions*

## 1.1 *Inner Working Angle (IWA) & Flux Ratio Detection Limit*

The IWA represents the minimum star-planet separation at which high-contrast observations can be conducted at the flux ratio detection limit. HabEx and LUVOIR are required to make 10σ detections and

---

spectrally resolved measurements of point sources $10^{10}$ times fainter than the central star, while, at the CBE level, CGI will be able to detect and spectrally characterize point sources only $10^{8}$ times fainter than the central star at 3-4$\lambda$/D in NFOV and SPEC modes. At this flux ratio, the CGI NFOV mode provides an IWA of 3$\lambda$/D for broad-band imaging around 575nm, and CGI SPEC provides spectroscopic characterization at an IWA of 3$\lambda$/D around 730 nm. In spite of the more aggressive flux ratio, the IWA for future missions is of the same order as CGI: 3.1 (HabEx), 3.5 (LUVOIR B) and 4 $\lambda$/D (LUVOIR A), all at visible wavelengths.

## 1.2 Visible Light Spectroscopic Capabilities

HabEx and LUVOIR have a broad spectral coverage that extends both bluer and redder than visible light. We limit the scope of this discussion to the 500-900 nm range that overlaps with CGI.

### 1.2.1 Spectral Bandwidth

For sensitivity reasons, and in order to limit the number of parallel channels, high-contrast coronagraphic measurements shall be performed over as wide a spectral range as possible. For HabEx and LUVOIR, the minimum instantaneous spectral bandwidth per coronagraph channel (or "mode") is set to 20%. In comparison, CGI will offer one 15% bandwidth spectroscopic mode (SPEC around 730 nm) and two 10% bandwidth broad-band imaging modes (NFOV around 575nm and WFOV around 825nm).

### 1.2.2 Spectral Resolution

For exo-Earth observations and identification of a potential oxygen A-band feature around 760nm, HabEx (LUVOIR) requires a minimum spectral resolution of 70 (140) in the visible. In comparison, CGI's SPEC mode provides a spectral resolution of 50 at 730 nm (varying from ~ 40 to ~65 over the 15% bandpass).

### 1.2.3 Multi-planet Spectroscopy

HabEx and LUVOIR both require an integral field spectrograph to obtain simultaneous spectra of multiple planets in a given exoplanetary system. CGI will use instead a slit + prism-based spectrograph, measuring the spectrum of a single off-axis point source at a time.

## 1.3 Polarimetric Capabilities

HabEx and LUVOIR are designed to obtain polarimetric measurements of exoplanets and debris disks. In particular, these missions will look at the linearly polarized signature of light reflected off surface oceans. CGI is expected to provide linear polarization fraction measurements of bright debris disks using two sets of Wollaston prisms to probe 4 different linear polarizations states (-45, 0, +45 and 90 deg). CGI's design specification is 3% systematic uncertainty (rms error) in linear polarized fraction.

## 1.4 Top-level CGI Design Specifications vs CBEs

If CGI performed strictly at its design specification level (no margin), its detection limit would be 5 times worse than at CBE, at an IWA degraded to 4$\lambda$/D. This is further away from the planet detection needs of future missions from a science yield standpoint. However, from a technology perspective, it still places very stringent requirements on lower level subsystems, in-line with the needs of future missions (see next section). CGI top-level spectral and polarimetric functionalities remain the same as at CBE, in family with the needs of future missions (except for the multi-object spectroscopic capability, already missing at CBE level).

## 2. CGI lower-level subsystems predicted performance and specifications vs future missions' needs

We focus here on the performance of the CGI/HabEx/LUVOIR lower-level subsystems (Table 2) needed to achieve the top-level capabilities discussed in the previous section. We concentrate on the following technologies: wavefront sensing and control, deformable mirrors and low flux detection.

| Lower Level Characteristics | CGI (CBEs) | CGI (Design Specifications) | HabEx Requirements | LUVOIR Requirements |
|---|---|---|---|---|
| **Wavefront Sensing and Control** | | | Assumes VVC6 | LUVOIR A: APLC; LUVOIR B: VVC6 |
| *Low Order Wavefront Sensing and Control* | Yes (up to Z11) | Yes (up to Z11) | Yes | Yes |
| *Pointing jitter after correction (mas rms per axis)* | 0.49 for V=5 | < 1.0 (NFOV) | < 0.3 | < 0.3 |
| *Differential pointing jitter (mas rms)* | 0.27 | 0.5 | < 0.3 | < 0.3 |
| *Residual Defocus drift (pm)* | 3 (temporal rms in OS9) | < 15 (NFOV) | < 1315 | A: < 33, B: < 7 |
| *Residual Astigmatism drift (pm)* | 3 temporal rms in OS9 | < 47 (NFOV) | < 157 | A: < 50, B: < 14 |
| *Residual Coma drift (pm)* | 2 (temporal rms in OS9) | < 4 (NFOV) | < 94 | A: < 1, B: < 8 |
| *Residual Spherical drift (pm)* | 2 (temporal rms in OS9) | < 2 (NFOV) | < 76 | A: < 2, B: < 4 |
| *High order wavefront drift in pm after any correction (weighted sum of all Zernikes with n+\|m\| >= 6)* | 5 (temporal rms in OS9) | < 50 | < 5 | < 5 |
| *Laser Metrology* | No | No | Yes M1-M2-M3 | Yes 6 per M1 segment-M2-M3 |
| **Deformable Mirrors (DM)** | | | | |
| *Number of actuators* | 48 x 48 | 48 x 48 | 64 x 64 | A: 128 x 128, B: 64 x 64 |
| *Number of DMs per coronagraph channel* | 2 | 2 | 2 | 2 |
| *DM stroke range ( μm)* | >0.5 | >0.5 | > 0.5 | > 0.5 |
| *DM stroke resolution (pm)* | 7.5 | <15 | 2.5 | 1.9 |
| **Low Flux Detection at 500 - 900nm** | | | | |
| *Photon counting* | Y | Y | Y | Y |
| *Detector Format* | 1024 x 1024 (imaging and spectroscopy) | 1024 x 1024 (imaging and spectroscopy) | 1024 x 1024 (imaging) 2048 x 2048 (IFS) | 1024 x 1024 (imaging) 4096 x 4096 (IFS) |
| *dQE at 550nm* | 0.50 | >0.47 | > 0.56 | 0.72 (CBE) |
| *Read-out-noise (amplified read noise divided by gain, in e- rms/pix /read)* | 0.015 | <0.02 | < 0.1 (CBE: 0) | 0 (CBE) |
| *Dark Current (e-/pix/s)* | 1.3 x 10-4 | <5.2 x 10-4 | < 4 x 10-4 (CBE: 3 x 10-5) | 3 x 10-5 (CBE) |
| *Clock-induced Charge (e-/pix/read)* | 5.00E-03 | <0.01 | < 6 x 10-2 (CBE: 1.3 x 10-3) | 1.3 x 10-3 (CBE) |
| *Lifetime at specified detector parameters* | 5 years | >21 months | >5 years | >5 years |
| *Minimum point source flux detectable per PSF core (e-/s)* | 0.14 (NFOV around 575 nm) 0.027 (SPEC around 730 nm) | <0.3 (NFOV around 575 nm) <0.04 (SPEC around 730 nm) | 0.01 (imaging) 0.0004 (IFS) | A: 0.06 (imaging) A: 0.0008 (IFS) |

***Table 2**: Comparison of CGI lower-level subsystem characteristics −CBEs as of August 2020 and design specifications − to the corresponding HabEx and LUVOIR requirements. For the CGI columns, the cells color coding indicates whether CGI performance CBE (or design specification) is better (green), in family with (blue) or significantly worse than (red) what is required for HabEx and the least stringent of LUVOIR A and B. The minimum detectable point source flux detectable by CGI per PSF core (last row) in spectroscopy (SPEC) mode is only comparable to the future missions needs for broad-band imaging, hence the red color coding. See text for details.*

## 2.1 Wavefront Sensing and Control

In comparing CGI's wavefront sensing and control capabilities to those required for future missions, we consider three types of wavefront instabilities: pointing jitter residuals, "low-order" and "high-order" aberrations drifts.

### 2.1.1 Pointing Control

*CGI performing at CBE*

For future missions, the pointing jitter must be maintained below 0.3 mas rms per axis, up to a stellar V magnitude of 6 (HabEx) and V=9 (LUVOIR). Using input line-of-sight (LoS) disturbances and flux levels consistent with the Roman Space Telescope in-flight expectations (~10 mas rms per axis), the CGI LoS sensing and control performance currently demonstrated in the lab – with both feedback and feedforward loops on – is 0.2 mas rms at V=2.5 and 0.35 mas rms at V=5 (Shi et al. 2018, Fig. 12). Using the LOWFS pointing feedback loop only, as is planned for the flight implementation, the residual LoS jitter is expected to increase to 0.49 mas rms for the worse axis. CGI's expected pointing control performance is then very much "in family" with the needs of future missions. Moreover, the HabEx and LUVOIR telescopes have much lower expected levels of LoS disturbances than the Roman Space Telescope (< 0.2 mas rms with HabEx micro-thrusters pointing system, < 0.3 mas rms with LUVOIR vibration isolation system) and higher pointing sensing at a given stellar magnitude because of their larger collecting area and (5 to 10x) higher core throughput[2]. Consequently, it could even be argued that the CGI LOWFS system is already sufficient to meet the HabEx and LUVOIR pointing control requirements.

*CGI performing at design specifications level*

CGI pointing jitter design specifications (1.0 mas rms) are set by the performance specification of the NFOV mode and are more relaxed than the 0.3 mas rms required for future missions.

### 2.1.2 Low-order aberrations

In addition to pointing jitter, CGI's LOWFS system will sense and control defocus, astigmatism, coma and spherical aberrations (as well as trefoil). These low-order aberrations contribute the lion's share of the wavefront drifts predicted by CGI structural thermal optical performance (STOP) models during representative observation sequences (Krist et al. 2018)

*CGI performing at CBE*

The CGI low-order aberration drift CBEs quoted in Table 2 (2nd column) are derived from the most recent CGI observing sequence ("OS9[3]") end-to-end STOP models and simulations of the observatory and CGI systems. They assume a Model Uncertainty Factor ("MUF") of 2 on the model-predicted observatory stability performance, and assume active correction of focus drifts with CGI LOWFS. All other low-order aberrations are predicted to be stable at the few pm level, at or below the LOWFS loop noise floor. Hence, active correction of these modes was not found to provide benefit in the OS9 CBE scenario.

CGI CBE values represent the expected temporal rms of each aberration over periods of tens of hours. These CGI values should be compared to the "not-to-exceed" aberration drifts required by future missions (Table 2, columns 4 & 5). For future missions, low-order aberrations stability requirements are computed assuming a nominal ("unperturbed") raw contrast of $10^{-10}$ at the required IWA. The maximum

---

[2] "Core throughput" is defined as the product of the occulter total transmission (any pupil or focal plane coronagraphic masks and downstream Lyot stop) by the fraction of the off-axis PSF falling within the PSF half-max contour.
[3] https://wfirst.ipac.caltech.edu/sims/Coronagraph_public_images.html

("not-to-exceed") drift allowed in each individual low-order aberration is defined as the amplitude causing a 1% *relative* degradation in raw contrast, i.e., yielding a net contrast increase of $10^{-12}$ at the IWA of each mission baseline coronagraph.

**Remarkably, the CGI CBEs for low-order aberration drifts are all significantly better than required for HabEx and LUVOIR B.** This important result has to do with the relative insensitivity to low-order aberrations of the Vector Vortex Charge 6 (VVC6) coronagraph currently baselined for HabEx and LUVOIR B. Indeed, the VVC6 is very resilient to defocus, astigmatism, coma and spherical, which are all part of the VVC6 coronagraph "null space" (Ruane et al. 2018, Juanola-Parramon et al. 2019, Pueyo et al. 2019). However, given the complexity of the Roman Space Telescope entrance pupil, a Vector Vortex Coronagraph is not easily adaptable and CGI uses Hybrid Lyot and Shaped Pupil Coronagraph masks (HLC and SPC) instead. CGI low-order wavefront stability CBEs are also better than required by the LUVOIR A APLC Coronagraph for defocus and astigmatism, and in family for coma and spherical.

*CGI performing at design specifications level*

As in the case of HabEx and LUVOIR, CGI low-order aberration stability design specifications correspond to a 1% relative contrast degradation per aberration. However, the initial contrast is set here to $3 \times 10^{-8}$ and at an IWA of $4\lambda/D$, in line with CGI baseline point-source detection and spectroscopy flux ratio specifications. The resulting CGI low-order aberration stability design specifications are given for defocus, astigmatism, coma and spherical (3rd column of Table 2) using the contrast degradation sensitivities of two CGI masks: HLC for broadband imaging around 575 nm (NFOV mode), and SPC ("bow-tie") for spectroscopic observations around 730 nm (SPEC mode).

As observed with CGI CBEs, CGI stability design specifications for defocus, astigmatism, coma and spherical aberrations remain more stringent than what is needed for HabEx. For CGI NFOV mode, all design specifications are in the few picometers to few tens of picometers range, i.e., either more stringent than or in family with the requirements of LUVOIR A/B, depending on the aberration considered.

**Strikingly, in order to reach a ~100x more modest raw contrast and contrast stability performance, CGI is designed to provide either better or similar *low-order* wavefront stability than future missions. Thus, while the heavily obscured Roman Space Telescope pupil does limit the ultimate contrast performance of CGI and its detection capabilities, it makes it an excellent technology demonstrator from a low-order wavefront control point of view.**

More generally, and as can be seen in Table 2, low-order wavefront stability requirements depend strongly on the telescope entrance pupil geometry – especially its degree of obscuration and segmentation – and on the coronagraph type used. The HabEx unobscured monolithic aperture design provides for instance more relaxed requirements than the LUVOIR segmented designs. This interplay between telescope and coronagraph designs is also evident when comparing the low-order wavefront stability requirements of the LUVOIR A (obscured and segmented aperture) APLC coronagraph and the LUVOIR B (unobscured and segmented aperture) VVC6 coronagraph.

### 2.1.3 High-order aberrations

We define "high-orders" here as all Zernike modes $Z_n^m$ with $n + |m| >= 6$; i.e.: all aberrations besides the ones considered in the previous section. CGI actively controls these modes only during periodic "touchups" on bright reference stars. CGI does not actively control these modes during observations of the target star; hence CGI high-order performance is limited by drift. Whether future missions will have to sense and actively correct for the drifts of such aberrations during science observations remains an open question, and one that CGI will help to answer. In any case, the final "high-order" wavefront drifts must be kept below a "not to-exceed" value of order 5pm for both the HabEx and LUVOIR mission concepts.

In comparison, the **CGI high-order wavefront drift CBEs predicted by the OS9 STOP model – with a MUF of 2 – show a temporal variation of 5 pm rms, in family with the requirements of future missions.**

Conversely to the case of low-order aberrations described above, these high-order aberrations are not rejected by the HabEx/LUVOIR baselined coronagraphs. Consequently, the CGI, HabEx and LUVOIR coronagraphs have similar sensitivities to these aberrations. Given that CGI's flux ratio specification is more modest than HabEx or LUVOIR's, CGI's high-order wavefront residuals specification is more relaxed than those of future missions, by a factor of ~10.

## 2.2 Deformable Mirrors (DMs)

The HabEx and LUVOIR B visible coronagraphs require a pair of DMs with at least 64x64 actuators and a minimum actuator stroke of 500 nm to create large full 360 deg broad-band dark holes for exoplanet searches. Due to its larger primary (15m), LUVOIR A requires an even larger format DM (128 x 128 actuators) to maintain a similar dark hole outer working angle.

In comparison, the CGI design is based on a pair of DMs with 48x48 actuators with a stroke of 500 nm or slightly higher. This is in-family with the needs of HabEx and LUVOIR B, but significantly smaller than the DM format required by LUVOIR A. At 7.5 pm CBE (<15 pm specification), the CGI DM stroke resolution (limited by its 16-bit DAC electronics) is significantly above the ~ 2 pm resolution required by future missions, which will require 18-bit resolution DAC electronics.

## 2.3 Low-flux Detection

CGI will provide the first in-space demonstration of photon-counting detection and spectroscopy of faint visible sources. CGI's expected Electron-Multiplying CCD (EMCCD) performance characteristics (QE, read-out noise, dark and clock-induced charge) and lifetime all meet the requirements of future missions' coronagraphs in the ~500-900 nm region.

For broad-band CGI imaging, the point-source flux levels expected to be detected (per PSF-core) are significantly higher than those required for future missions. It is only for CGI *spectroscopic* (R~50) observations that CGI detected fluxes become comparable to the *broad-band imaging* sensitivity needs of future missions. In order to spectrally characterize exo-Earths, future missions need to detect fluxes 10 to 100 times fainter than what CGI is required to do. This does not call for better detectors in the 500-900 nm region, but instead for the larger collecting area and higher core throughput provided by the HabEx and LUVOIR coronagraphic architectures. We also note that future missions will need higher QE in the red visible part (> 900 nm) than provided by CGI's EMCCDs.

**Even if CGI only met its detector performance specifications on QE and noise properties rather than the current CBEs, it would still meet all the visible (500-900 nm) detector requirements of future missions**, except for lifetime (21 months vs 5 years). The faintest signals detected via CGI spectroscopic measurements would still be in family with those required for broad-band imaging with future missions.

# 3. Summary

We compared CGI current best estimates ("CBEs") of performance – as well as CGI specifications – to the needs of possible future exoplanet missions such as HabEx and LUVOIR. This analysis is by no means exhaustive, but key subsystems and functionalities were considered.

**The main take-away message is that while the detection limits of CGI and future missions differ by ~100x , the performance of many of their respective subsystems are very much in family, making CGI an excellent technology demonstration for these future missions. In fact, in many key technical areas, such as pointing jitter control, low-order wavefront control and detector properties, CGI has to work as well or better than both HabEx and LUVOIR.** This result has to do with the heavily obscured entrance pupil of the Roman Space Telescope, which makes CGI more sensitive to common optical aberrations and also limits its off-axis throughput.

| Parameter | CGI vs. Future Missions (FM) with unobscured apertures: HabEx & LUVOIR B |
|---|---|
| $10\sigma$ Flux Ratio Detectable at $3\,\lambda/D$ | $10^{-8}$ vs $10^{-10}$ <br> Roman Space telescope Pupil is challenging |
|  |  |
| Wavefront error sources | Comparable <br> Phase & "new physics" (amplitude and polarization) |
| Pointing Jitter Control | Comparable <br> CGI lab: ~ 0.35 mas rms V=5 star, CBE = 0.49 mas rms <br> FM: 0.3mas NTE* |
| Low-order Control | ~ 100x better than required for HabEx <br> better than LUVOIR needs <br> Complex Roman Space Telescope pupil: <br> trading low-order sensitivity for overall throughput |
| High-order drift | Comparable (~5pm) <br> CGI: $1\sigma$ prediction** <br> FM: NTE* |
| # of DMs | Same (2) |
| DM Stroke Resolution | ~4X worse (7.5pm vs 2pm) <br> Engineering problem, not physics problem |
| DM Actuator Count | 48x48 vs 64x64 |
| EMCCD | Comparable at V-band <br> Bit better: dark current, clock-induced charge <br> Bit worse: QE at UV/red at 5 years (rad hard) |

***Table 3***: *CGI CBE performance compares favorably to future missions' (FM) requirements. (\*): NTE = Not-to-exceed value. (\*\*): Predicted performance from ROMAN Space Telescope end-to-end STOP model, with MUF=2 on observatory model.*

At CGI CBE level of performance (Table 3), we find that, for instance:

- o CGI pointing control residuals are comparable to the needs of future missions,
- o The predicted level of residual low-order wavefront drifts is ~10 to 100x better than required for HabEx, and is either better than or in family with the LUVOIR requirements.
- o The predicted level of higher-order wavefront drifts is in family with what is required for future missions
- o The DM characteristics (format) and predicted performance are in family with the needs of HabEx and LUVOIR B, except in terms of stroke resolution, which can be remedied with higher-precision DAC electronics.
- o The CGI photon-counting EMCCD performance characteristics (noise and QE) are better than what is needed for future missions at visible wavelengths, and their lifetime is comparable.

All of these conclusions still hold in the case where CGI only operates at its design specification level, with 2 exceptions:

- o CGI high-order wavefront drift design specifications are 10x looser than required for future missions. Nevertheless, CGI observations will be very informative in this respect, by comparing in-flight deformation levels to those predicted by the STOP models. This will help establish whether higher-order active wavefront control and extensive metrology are needed to control higher orders to the levels required by future missions.
- o The CGI EMCCDs required lifetime at specified performance characteristics is > 21 months, to be compared to > 5 years for future missions.

Finally, while this analysis demonstrates the relevance of CGI's main subsystems, the unique value of CGI to future missions resides in its system-level demonstration of active coronagraphy in space, allowing us to understand for the first time how all these subsystems interact together and with the overall observatory in the space environment.

### *Acknowledgements*

Part of the research was carried out at the Jet Propulsion Laboratory, California Institute of Technology, under a contract with the National Aeronautics and Space Administration. © 2020 California Institute of Technology. Government sponsorship acknowledged.